\documentclass[
reprint,
superscriptaddress,
amsmath,amssymb,
aip,
apl,
]{revtex4-1}

\usepackage[utf8]{inputenc}
\usepackage{graphicx,import} 
\usepackage{verbatim}
\usepackage{dcolumn}
\usepackage{bm}
\usepackage[colorlinks]{hyperref}
\usepackage{xcolor}

\definecolor{bluegray}{RGB}{40,180,160}
\definecolor{navygray}{RGB}{110,140,170}

\definecolor{meadowgreen}{RGB}
{0,128,0}
\definecolor{coolbrown}{RGB}
{165,42,42}

\hypersetup{
citecolor={bluegray}, 
linkcolor={navygray},
}

\usepackage{braket}
\usepackage{amsmath}
\usepackage{amssymb}
\usepackage{SIunits}
\usepackage{changes}
\usepackage{lipsum}
\usepackage{textcomp}
\usepackage{verbatim}
\usepackage{chngcntr} 

\usepackage[colorinlistoftodos,prependcaption,textsize=tiny]{todonotes}
\usepackage{multirow}
\usepackage{transparent}

\DeclareUnicodeCharacter{2212}{-}
\newcommand{\be}{\begin{equation}}
\newcommand{\ee}{\end{equation}}
\newcommand{\ba}{\begin{equation}}
\newcommand{\ea}{\end{equation}}
\newcommand{\bea}{\begin{eqnarray}}
\newcommand{\eea}{\end{eqnarray}}

\newcommand{\eref}[1]{Eq.~(\ref{#1})}

\newcommand{\rref}[1]{(\ref{#1})}

\usepackage{amssymb,amsmath,amsthm,enumitem}
\begin{document}

\title{Phonon traps reduce the quasiparticle density in superconducting circuits}

\author{Fabio Henriques}
\thanks{ \color{meadowgreen} Both authors contributed equally}
\affiliation{PHI, Karlsruhe Institute of Technology, 76131 Karlsruhe, Germany}

\author{Francesco Valenti}
\thanks{ \color{meadowgreen} Both authors contributed equally}
\affiliation{PHI, Karlsruhe Institute of Technology, 76131 Karlsruhe, Germany}
\affiliation{ \mbox {IPE,~Karlsruhe Institute of Technology}, 76344 Eggenstein-Leopoldshafen, Germany}

\author{Thibault Charpentier}
\affiliation{PHI, Karlsruhe Institute of Technology, 76131 Karlsruhe, Germany}

\author{Marc Lagoin}
\affiliation{PHI, Karlsruhe Institute of Technology, 76131 Karlsruhe, Germany}

\author{Clement Gouriou}
\affiliation{PHI, Karlsruhe Institute of Technology, 76131 Karlsruhe, Germany}

\author{Maria Mart{\'i}nez}
\affiliation{LFNAE, Universidad de Zaragoza, 50009 Zaragoza, Spain}

\author{Laura Cardani}
\affiliation{INFN-Sezione di Roma, Piazzale Aldo Moro 2, 00185 Roma, Italy}

\author{Marco Vignati}
\affiliation{INFN-Sezione di Roma, Piazzale Aldo Moro 2, 00185 Roma, Italy}

\author{Lukas Gr{\"u}nhaupt}
\affiliation{PHI, Karlsruhe Institute of Technology, 76131 Karlsruhe, Germany}

\author{Daria Gusenkova}
\affiliation{PHI, Karlsruhe Institute of Technology, 76131 Karlsruhe, Germany}

\author{Julian Ferrero}
\affiliation{PHI, Karlsruhe Institute of Technology, 76131 Karlsruhe, Germany}

\author{Sebastian T. Skacel}
\affiliation{PHI, Karlsruhe Institute of Technology, 76131 Karlsruhe, Germany}

\author{Wolfgang Wernsdorfer}
\affiliation{PHI, Karlsruhe Institute of Technology, 76131 Karlsruhe, Germany}
\affiliation{INT, Karlsruhe Institute of Technology, 76344 Eggenstein-Leopoldshafen, Germany}

\author{Alexey V. Ustinov}
\affiliation{PHI, Karlsruhe Institute of Technology, 76131 Karlsruhe, Germany}
\affiliation{\mbox{RQC, National University of Science and Technology MISIS, 119049 Moscow, Russia}}

\author{Gianluigi Catelani}
\affiliation{ \mbox{JARA Institute for Quantum Information (PGI-11),~Forschungszentrum J{\"u}lich, 52425 J{\"u}lich, Germany}}

\author{Oliver Sander}
\affiliation{ \mbox {IPE,~Karlsruhe Institute of Technology}, 76344 Eggenstein-Leopoldshafen, Germany}

\author{\mbox{Ioan M. Pop}}
\email{ioan.pop@kit.edu}
\affiliation{PHI, Karlsruhe Institute of Technology, 76131 Karlsruhe, Germany}
\affiliation{INT, Karlsruhe Institute of Technology, 76344 Eggenstein-Leopoldshafen, Germany}

\date{\today}

\begin{abstract}
Out of equilibrium quasiparticles (QPs) are one of the main sources of decoherence in superconducting quantum circuits, and are particularly detrimental in devices with high kinetic inductance, such as high impedance resonators, qubits, and detectors. Despite significant progress in the understanding of QP dynamics, pinpointing their origin and decreasing their density remain outstanding tasks. The cyclic process of recombination and generation of QPs implies the exchange of phonons between the superconducting thin film and the underlying substrate. Reducing the number of substrate phonons with frequencies exceeding the spectral gap of the superconductor should result in a reduction of QPs. Indeed, we demonstrate that surrounding high impedance resonators made of granular aluminum (grAl) with lower gapped thin film aluminum islands increases the internal quality factors of the resonators in the single photon regime, suppresses the noise, and reduces the rate of observed QP bursts. The aluminum islands are positioned far enough from the resonators to be electromagnetically decoupled, thus not changing the resonator frequency, nor the loading. We therefore attribute the improvements observed in grAl resonators to phonon trapping at frequencies close to the spectral gap of aluminum, well below the grAl gap.
\end{abstract}

\maketitle

Superconducting circuits play a central role in a variety of research and application areas, such as solid state quantum optics \cite{gu2017microwave}, metrology \cite{taylor1989new, pekola2013single}, and low temperature detectors \cite{day2003broadband, zmuidzinas2012superconducting}. In particular, the field of superconducting qubits has grown impressively during the last decade \cite{gambetta2017building, krantz2019guide}. In these devices quantum states can live for up to tens of microseconds, while gate times can be as short as tens of nanoseconds  \cite{yan2016the, rol2017restless, klimov2018fluctuations, bronn2018high}. Nevertheless, coherence times need to be further improved by orders of magnitude in order to be able to perform quantum error correction \cite{fowler2012towards, ofek2016extending} with an affordable hardware overhead.

\begin{figure}[h!]
\begin{center}
\vspace{1 mm}
\def\svgwidth{1 \columnwidth}  
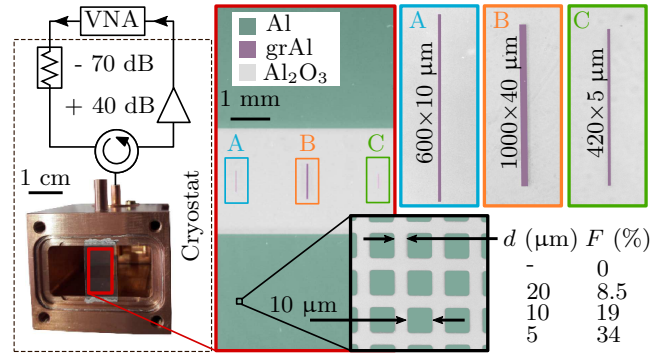
\caption{Photograph of a Cu waveguide housing a $15\times 8 \times 0.33 $~mm$^3$ sapphire chip (red box) supporting three $20$~nm thick grAl resonators (highlighted in magenta, labeled A-C), and a square lattice of $60$~nm thick Al phonon traps (jade-colored squares outlined in the black box), all patterned via optical lift-off lithography. We sweep the spacing $d$ between Al islands to obtain different phonon trap filling factors $F$, defined as the fraction of substrate covered by traps, and listed in the inset table. The chip is glued to the waveguide using silver paste. The waveguide, anchored to the mixing chamber of a dilution cryostat at $15$~mK, is connected to a reduced schematics of the microwave reflection measurement setup (cf. Suppl. Mat.~\ref{app_cryo}). }
\label{fig_samples}
\end{center}
\end{figure}

One of the main sources of decoherence in superconducting devices at millikelvin temperatures are out of equilibrium quasiparticles (QPs) \cite{aumentado2004nonequilibrium, barends2008quasiparticle, shaw2008kinetics, catelani2011quasiparticle, de2011number, riste2013millisecond, wang2014measurement, nsanzineza2014trapping, serniak2018hot}, which can be viewed as broken Cooper pairs (CPs). Quasiparticles can be particularly damaging in high kinetic inductance circuits \cite{vool2014non, janvier2015coherent, gustavsson2016suppressing, grunhaupt2018loss, grunhaupt2018granular}, which are a promising avenue for protected qubits \cite{groszkowski2018coherence} and hybrid superconducting-semiconducting devices \cite{viennot2015coherent, mi2017strong, landig2018coherent}. Proposed mechanisms for CP breaking include stray infrared radiation \cite{barends2011minimizing,houzet2019photon}, direct microwave drive \cite{de2014evidence, patel2018phonon}, and high energy phonons in the device substrate created by environmental or cosmic radioactivity \cite{swenson2010high, bespalov2016theoretical, karatsu2019mitigation}.  The latter is particularly damaging because it gives rise to correlated QP bursts in multiple devices on the same chip \cite{swenson2010high, moore2012position}, possibly resulting in correlated errors, further complicating the task of error correction.

Quasiparticle mitigation strategies can be divided into two categories. One approach consists of removing QPs, e.g. by trapping them in normal metals \cite{riwar2016normal,hosseinkhani2017optimal} and vortices \cite{nsanzineza2014trapping, wang2014measurement, taupin2016tunable}, or by pumping them outside the susceptible region of the circuit \cite{gustavsson2016suppressing}. The other approach consists of preventing CP breaking, e.g. by filtering and shielding from radiation with frequency above the superconducting spectral gap \cite{barends2011minimizing}.

In this Letter we describe a complementary method of preventing CP breaking through the reduction of high energy phonons in the substrate. We demonstrate that the figures of merit of superconducting grAl resonators, such as single photon internal quality factor and noise spectral density, can be improved by surrounding them with lower gapped islands made of pure aluminum, which act as phonon traps.

Our approach is similar to that of Refs.~\cite{karatsu2019mitigation,valenti2018interplay}, in which it has recently been demonstrated that surrounding kinetic inductance detectors with a lower gapped superconducting film reduces the number of measurable QP bursts by an order of magnitude and the noise equivalent power by a factor three. Phonon traps downconvert the frequency of high energy substrate phonons to that of their own spectral gap via inelastic electron-phonon interactions. Therefore, phonons resulting from recombination in the traps are unable to break CPs in the circuit. 

This phenomenological model is detailed in Ref.~\cite{valenti2018interplay}, where it is shown that the phonon traps’ efficiency increases with the difference between the spectral gaps of the circuit and trap materials. In the following, we demonstrate that the efficiency rapidly scales with the traps' surface. We report more than a factor two improvement in single photon internal quality factors, as well as a reduction of the noise amplitude by an order of magnitude, for traps covering as little as a third of one side of the substrate.

We use superconducting grAl resonators because they can have a kinetic inductance fraction close to unity, thus providing a high susceptibility to QPs, while also retaining high internal quality factors $Q_i$ in the range of $10^4-10^5$ in the single photon regime \cite{rotzinger2016aluminium, grunhaupt2018loss, zhang2019microresonators}. Granular aluminum is a composite material made of self-assembled Al grains, $3-4$ nm in diameter, embedded into an amorphous AlO$_\text{x}$ matrix \cite{cohen1968superconductivity,deutscher1973granular}. The thickness of the oxide shells can be tuned by the oxygen pressure during thin film deposition, which results in resistivities spanning $\rho = 1-10^4 \; \upmu \Omega \; $~cm. Thanks to the corresponding kinetic inductances up to the nH$/\Box$ range, grAl has recently attracted interest as a material for high impedance quantum circuits \cite{grunhaupt2018granular}. The superconducting gap of grAl is dome-shaped as a function of resistivity \cite{pracht2016enhanced, levybertrand2019electrodynamics}; for films grown on substrates at room temperature, the critical temperature has a maximum $T_{c, \text{ max}} \approx 2.1$~K for $\rho \approx 400 \; \upmu \Omega $~cm \cite{levybertrand2019electrodynamics}, significantly above the critical temperature of thin film aluminum, $T_{c,\text{ Al}} \approx 1.4$~K. The resulting difference in the spectral gaps allows Al to be used as a phonon trap for grAl circuits \cite{valenti2018interplay}.

We fabricate all resonators in the same lithography step using optical lift-off and electron beam evaporation of a $20$ nm thick grAl film on a $330 \; \upmu$m thick, double-side polished $c$-plane sapphire wafer. We employ a grAl film with $\rho = 5\; \text{m} \Omega  $~cm and $T_c\approx1.8$~K in order to maximize the sensitivity to QP bursts, while remaining a factor two below the edge of the superconductor-to-insulator transition \cite{levybertrand2019electrodynamics}. The resulting kinetic inductance per square is $L_K = 2 \; \text{nH}/\Box$, orders of magnitude larger than the geometric inductance. 

As shown in Fig.~\ref{fig_samples}, and similar to Ref.~\cite{grunhaupt2018loss}, each chip hosts three grAl resonators with sizes of $600\times 10$, $1000\times 40$, and $420\times 5 \; \upmu$m$^2$, which we label A, B and C, respectively. The resonators are surrounded by a square lattice of $10 \times 10 \; \upmu$m$^2$ aluminum islands, $60$ nm thick, deposited in a second lithographic step, using the same lift-off technique employed for the resonators. We fabricate three types of chips with various lattice parameters, $d=20$, $10$, and $5\; \upmu$m, in order to achieve an increasingly larger phonon trap filling factor ($F$) of $8.5$, $19$, and $34\%$, respectively. We also fabricate a witness chip without traps ($F=0$). 

Using the phonon trapping model of Ref. \cite{valenti2018interplay} for grAl resonators in the presence of Al islands (cf. Suppl. Mat. \ref{app_model}) we show that the internal dissipation rate $1/Q_i$ and QP burst rate $\Gamma_B$ decrease as a function of increasing $F$:\begin{align}  
 \frac{1}{Q_i} & = \frac{1}{Q_{i,0}} \sqrt{1+( \beta  F)^2 - \sqrt{2( \beta  F)^2 + ( \beta F)^4}}  \label{Q_fit} \\
\Gamma _B & =  \Gamma_0 \frac{ \Lambda } {F +  \Lambda } \label{gammab_fit} . 
\end{align}Here, $Q_{i,0}$ and $\Gamma_0$ are the internal quality factor and QP burst rate for $F=0$ respectively. The coefficient $\beta$ is a phenomenological constant which accounts for the rates of phonon generation, scattering and thermalization, and $\Lambda $ is the ratio between the rates of phonon thermalization to the sample holder and phonon absorbption in the traps.

We would like to note that both the island size and the lattice parameter $d$ are two orders of magnitude larger than the wavelength of phonons resulting from QP recombination in grAl and Al, which is in the range of $50-100$~nm, considering a speed of sound in sapphire of about $10$~km/s \cite{winey2001r}. The propagation of phonons in the substrate is thus unhindered by gaps in the phonon dispersion relation. However, phononic crystal engineering could also be a viable phonon mitigation approach, as demonstrated by the shielding of optomechanical resonators from phonons at GHz frequencies \cite{chan2011laser}.

 The sapphire chip is glued to the Cu waveguide shown in Fig.~\ref{fig_samples} using silver paste. We use thin indium wire to ensure both tight sealing and electrical contact between the waveguide and its cap (not shown). The resonators couple to the TE10 waveguide mode, providing a low loss microwave environment \cite{kou2018simultaneous}. The waveguide is placed into successive thermal and magnetic shields, and the microwave lines are heavily attenuated and shielded, similar to the setup of Ref.~\cite{grunhaupt2018loss}.

\begin{figure}[!t]
\begin{center}
\def\svgwidth{1 \columnwidth}  
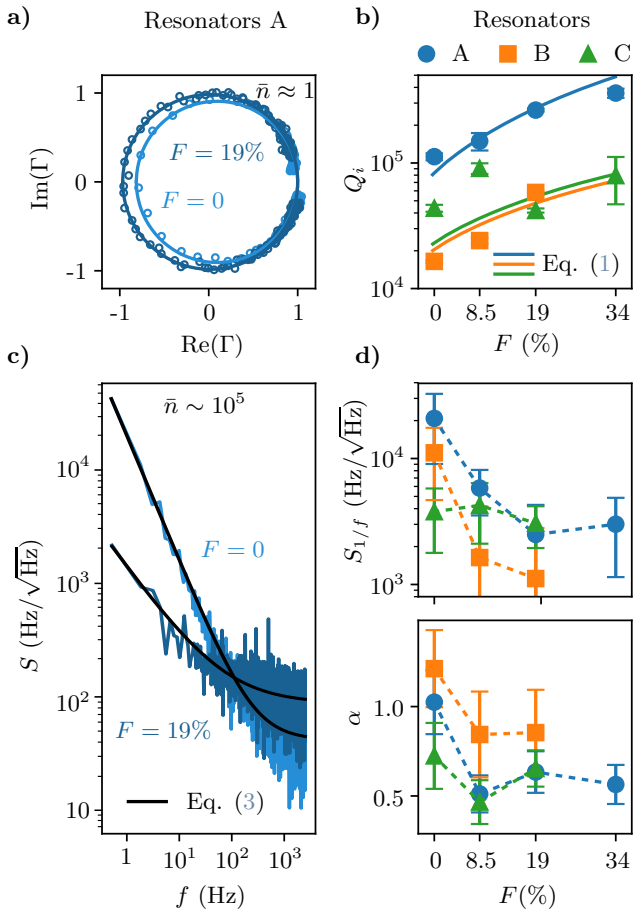
\caption{Effect of phonon trapping on resonator dissipation and noise. \textbf{a)} Typical measurement of the real and imaginary components of the reflection coefficient $\Gamma$, normalized to the waveguide response, for two resonators A with different filling factors $F$~$= 0$ (light blue) and  $F$~$= 19\%$ (blue), in the single photon regime. \textbf{b)} Internal quality factors in the single photon regime extracted from the circle fit (cf. Suppl. Mat. \ref{app_circlefit}), plotted as a function of the filling factor. Error bars represent the fitting routine uncertainty. The solid lines are fits to Eq.~\eqref{Q_fit}. \textbf{c)} Noise spectral density $S(f)$ of the two resonators shown in panel~\textbf{a}. We compute the spectra from time traces, each four seconds long, in which no QP bursts are present (in contrast with time traces shown in Fig.~\ref{fig_events}a). We fit the data (black lines) with the phenomenological model of Eq.~\eqref{noise_eq}. \textbf{d)} Fitted noise amplitude $S_{1/f}$ (top panel) and exponent $\alpha$ (bottom panel) plotted as a function of the filling factor. We obtain the plotted values by averaging over tens of spectra. Error bars represent one standard deviation. The dashed lines connecting the markers are guides to the eye.}
\label{fig_parameters}
\end{center}
\end{figure}

\begin{figure*}[th!]
\begin{center}
\def\svgwidth{1 \textwidth}  
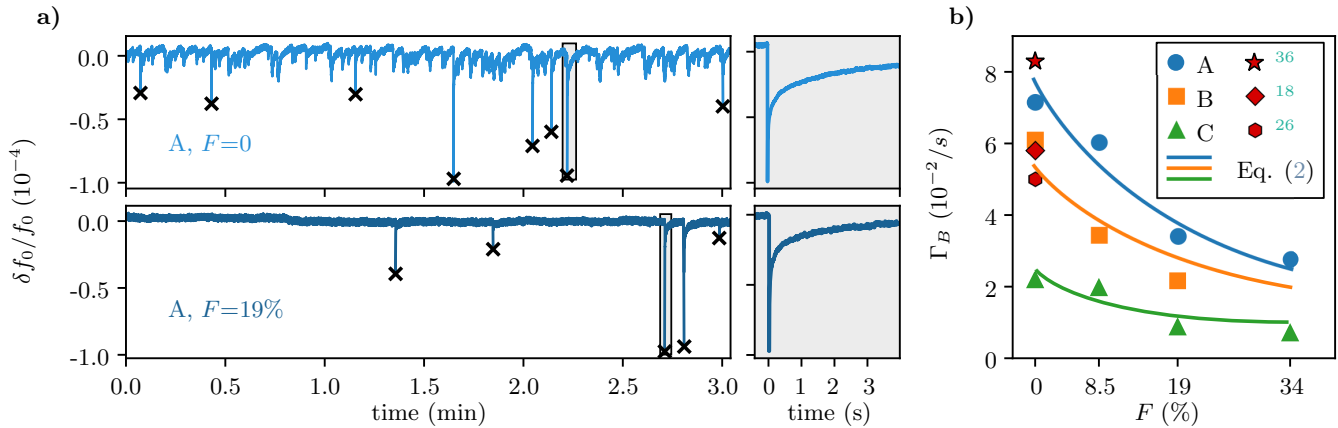
\caption{Effect of phonon trapping on QP bursts. \textbf{a)} Typical time trace measurement of $\delta f_0/f_0$ for resonators A, filling factor $F$~$=0$ (top panel, ligth blue) and $F=19\%$ (bottom panel, blue), with zoom-ins on single QP bursts. Quasiparticle bursts (see main text for a discussion on their possible origins) are marked with a black cross. Increasing the filling factor $F$ yields a twofold improvement, reducing both the low frequency baseline noise (cf. Fig.~\ref{fig_parameters}c and d) and the number of observed bursts. In the right hand panels we plot the typical frequency evolution during the first few seconds after a typical QP burst (black boxes in the respective left hand panel). The relaxation after the burst consists of an initial steep decay, followed by an exponential tail. Note that increasing the filling factor of the traps does not change the QP relaxation dynamics in the grAl resonators. \textbf{b)}~Measured rate of QP bursts $\Gamma_B$ for all resonators and all filling factors, averaged over ten hours (cf. Suppl. Mat.~\ref{app_spectrograms}). The QP burst rate decreases monotonically with the filling factor for all resonators. The solid lines are fits to Eq.~\eqref{gammab_fit}. For comparison, we also show previously reported QP burst rates (red markers) for both grAl \cite{grunhaupt2018loss} and Al \cite{swenson2010high,de2011number}.
}
\label{fig_events}
\end{center}
\end{figure*}

 We measure the complex reflection coefficient $\Gamma$ of the grAl resonators using a Vector Network Analyzer (VNA). In Fig.~\ref{fig_parameters}a we plot typical measurement results of $\Gamma$ vs. frequency in the complex plane for two resonators of type A, with different filling factors, $F=0$ and $F=19\%$. The larger diameter of the reflection circle for the resonator surrounded by phonon traps indicates reduced losses. 
 
 In Fig.~\ref{fig_parameters}b we plot the fitted internal quality factors $Q_i$ as a function of the phonon trap filling factor for all resonators. We observe an overall trend of $Q_i$ increasing with $F$, which can be fitted to Eq. \eqref{Q_fit} by choosing $\beta=9$ for all resonators (cf. Suppl. Mat. \ref{app_model}). This allows us to extrapolate that going from $F=0$ to $F \to 1$ the single photon $Q_i$ can be in principle increased by up to an order of magnitude.
 
 The measurements of the internal quality factors presented in Fig.~\ref{fig_parameters}a and b were obtained using a readout drive corresponding to a circulating photon number $\bar{n} \approx 1$. The photon number is calibrated using the formula $\bar{n} = 4P_\text{cold} Q_\text{tot}^2/\hbar \omega _0^2Q_c$, where $P_\text{cold}$ is the on-sample drive, $Q_\text{tot}^{-1} = Q_i^{-1} + Q_c ^{-1}$ and $Q_c$ is the coupling quality factor. At stronger drives ($\bar{n} \gg 1$) $Q_i$ is further increased, either by saturating dielectric loss \cite{hunklinger1972saturation,golding1973nonlinear} or enhancing QP diffusion \cite{levenson2014single,gustavsson2016suppressing}; however, it is also less correlated with $F$ (cf. Suppl. Mat.~\ref{app_qinbar}), possibly due to the onset of more complex QP and phonon dynamics.  Resonant frequencies and coupling quality factors for all resonators are summarized in Suppl. Mat.~\ref{app_circlefit}. Resonator B with $F=34\%$ could not be measured, most likely due to its resonant frequency being outside of the frequency band of the setup.

In Fig.~\ref{fig_parameters}c we plot the noise spectral density $S(f)$ for resonator A with filling factors $F$~$=0$ and $19\%$ measured at the highest power before bifurcation ($\bar{n}\sim 10^5$). Note the order of magnitude reduction in low frequency noise amplitude for the sample with phonon traps; the noise floor at high frequency is given by the readout electronics. Interestingly, the amplitude of the $1/f$ noise does not depend on $\bar{n}$ (cf. Suppl. Mat.~ \ref{app_nsdpow}). We fit the noise spectra with the phenomenological model 
\begin{equation} \label{noise_eq}
S(f) = S_0 + \frac{S_{1/f}}{(f/\mathrm{1\,Hz})^\alpha}.
\end{equation}
In Fig.~\ref{fig_parameters}d we plot the fitted amplitude $S_{1/f}$ and exponent $\alpha$ for noise spectra of all measured resonators vs. $F$. We observe an overall decreasing trend for both amplitude and exponent of the noise as a function of the phonon trap filling factor. This trend is consistent with the observed increase in $Q_i$ (cf. Fig.~\ref{fig_parameters}b), and indicates QP generation-recombination as a primary source of noise~\cite{de2011number}.

In Fig.~\ref{fig_events}a we show the time evolution of the resonant frequency $f_0$ for resonators A with $F$~$=0$ and $19\%$. The time traces of the resonant frequency show noise that is qualitatively similar to the one reported in Refs.~\cite{swenson2010high,de2011number,grunhaupt2018loss}: stochastic QP bursts, which abruptly lower the resonant frequency and are followed by a relaxation tail, interspersed on top of a background of fluctuations. The resonator surrounded by phonon traps with $F=19\%$ shows a reduction in both the fluctuations and the number of measured QP bursts, indicating a significant reduction in non-equilibrium phonons with energies above the spectral gap of grAl. The QP relaxation after a burst (cf. right hand panels of Fig.~\ref{fig_events}a) is unaffected by $F$. We fit the exponential tails with the same methodology of Ref.~\cite{grunhaupt2018loss} and obtain the QP lifetime $\tau_\text{qp} = 0.5 \pm 0.1$~s for all resonators and $F$ values.

We measure a constant QP burst rate over the course of several days (cf. Suppl. Mat.~\ref{app_days}). In Fig.~\ref{fig_events}b we show that the rate of bursts decreases monotonically with $F$ for all resonators. We interpret this as a decrease in the probability that pair-breaking phonons reach the resonators; the larger the filling factor, the more effective the phonon trapping. We fit the QP burst rate to Eq. \eqref{gammab_fit} using the phonon relaxation ratio $\Lambda=0.18$ for all resonators (cf. Suppl. Mat. \ref{app_model}), which shows that the QP burst rate can in principle be reduced by a factor $(1+\Lambda)/\Lambda \sim 6$ for $F \to 1$. We would like to note that for similarly sized substrates without phonon traps the measured QP burst rates are comparable (cf. Fig.~\ref{fig_events}b and Refs.~\cite{swenson2010high,de2011number,grunhaupt2018loss}). The impact rate of cosmic muons on the substrate can account for up to $30\%$ of the measured rate \cite{barnett1996review}, with the rest possibly originating from various environmental radioactive sources, which should be further investigated.

In summary, we measured 11 granular aluminum  resonators with a resistivity $\rho \approx 5 \; \text{m} \Omega$~cm, corresponding to a kinetic inductance of $2 \; \text{nH}/ \Box$. Out of these, 8 were fabricated on chips containing aluminum islands with varying filling factor. The aluminum islands are electromagnetically decoupled from the resonators, and act as phonon traps due to their lower superconducting gap compared to grAl. When increasing the density of phonon traps, we observe three types of improvement of resonator performance: internal quality factors in the single photon regime increase by up to a factor three, the $1/f$ noise is reduced by an order of magnitude, and the rate of QP bursts is halved. These results indicate that non thermal phonons in the substrate play an important role in the generation of non-equilibrium QPs in superconducting circuits, and phonon frequency down-conversion can be a remarkably effective strategy to reduce QP density.

Future work should focus on maximizing the filling factor $F$ and decreasing the phonon relaxation ratio $\Lambda$ by employing traps with decreased gap and increased thickness. Further improvements might be achieved by engineering the phonon dispersion relation in the substrate, by placing the superconducting devices in regions of lower phonon density, by identifying and removing hot phonon sources, or by improving phonon thermalization, thus reducing the QP burst rate $\Gamma_0$.

See supplementary material for information on the cryogenic setup, the phonon trapping model of Ref.~\cite{valenti2018interplay}, the power dependence of resonator noise and dissipation, and the measured QP bursts over longer timespans.

We are grateful to A. Monfardini, J. Baselmans and P. de Visser for insightful discussions, and to L. Radtke and A. Lukashenko for technical support. Facilities use was supported by the KIT Nanostructure Service Laboratory (NSL). Funding was provided by the Alexander von Humboldt foundation in the framework of a Sofja Kovalevskaja award endowed by the German Federal Ministry of Education and Research, and by the Initiative and Networking Fund of the Helmholtz Association, within the Helmholtz Future Project Scalable solid state quantum computing. This work was partially supported by the Ministry of Education and Science of the Russian Federation in the framework of the Program to Increase Competitiveness of the NUST MISIS, contracts no. K2-2016-063 and K2-2017-081.

\bibliography{refe}

\counterwithin{figure}{section}
\counterwithin{equation}{section}

\onecolumngrid
\newpage

\section*{Supplementary Material}
\hrulefill

\vspace{2 cm}

\section{Measurement setup}\label{app_cryo}

\begin{figure*}[h!]
\begin{center}
\def\svgwidth{1 \textwidth}  
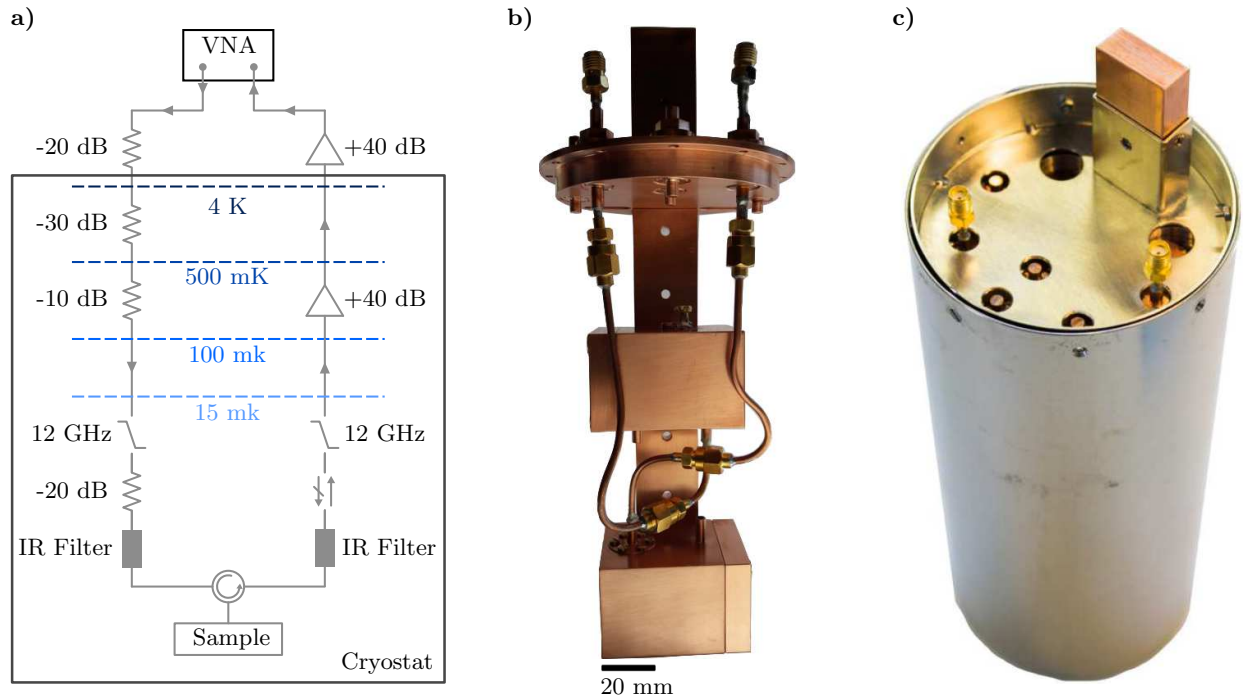
\caption{Schematics of the measurement setup. \textbf{a)} Diagram of the dilution cryostat and the microwave lines. \textbf{b)} A Cu rod hosting two mounted waveguides. \textbf{c)} Photograph of the rod inserted in the barrel. Only the outer $\upmu$-metal  shield is visible.}\label{cryo}
\end{center}
\end{figure*}

 In Fig~\ref{cryo}a we show a schematic of the employed cryogenics setup, together with its microwave lines. We use circulators in order to be able to perform reflection measurements with separate input and output lines. Attenuators and IR filters are used along the input line throughout the different temperature stages to thermalize the input rf field. The reflected signal is retrieved by both a cold high electron mobility transistor and a room temperature amplifier. Back-propagating noise is reduced with the use of an isolator. The Cu waveguides containing the chips are mounted on Cu rods as shown in Fig~\ref{cryo}b, which are in turn inserted into shielding cylinders (Fig~\ref{cryo}c), composed of an outer $\upmu$-metal layer and inner Cu/Al bilayer, shielding from stray magnetic and IR fields, respectively.
\vspace{10 mm}
\vspace{10 mm}

\newpage 
\section{Phonon trapping model}\label{app_model}

In the following we review the model introduced in Ref.~\cite{valenti2018interplay}, and apply it in order to quantify the effect of the phonon traps.

\subsection{Quality factor}

Due to the high impedance of the grAl films, we posit QPs to be the dominating source of losses. Under this assumption, the internal quality factor $Q_i$ is proportional to the quasiparticle (areal) density,
\be
\frac{1}{Q_i} = c \frac{N_G}{A_G} \label{Qeq},
\ee
where parameters such as thickness, as well as unknown ones, are absorbed into the coefficient $c$. With $N_\alpha$ we denote the number of QPs in the film of area $A_\alpha$, either a grAl resonator ($\alpha = G$) or the phonon traps lattice ($\alpha = A$). We task ourselves with finding how the QP density depends on the amount of phonon trapping surface.

We model the dynamics of ``hot'' phonons (i.e., able to break CPs in $G$) and QPs in a phenomenological way, with rate equations of the Rothwarf-Taylor type. For QPs in $G$, the relevant processes are generation from pair breaking by hot phonons (rate $b_G$) and recombination (rate $r_G$). Similarly, for QPs in $A$ we have generation by pair breaking ($b_A)$ and recombination ($r_A)$, but also scattering to lower energies (rate $s_A$). For the phonons, we assume some generation mechanism with rate $g_P$, in addition to generation/recombination in both $G$ and $A$. Finally, we include also the possibility that phonons escape from substrate or otherwise decay with a rate $e_P$; note that this parameter is the only addition to the model previously developed in Ref.~\cite{valenti2018interplay}. The rate equations are then:
\bea
\dot{N}_G & = & -2r_G N_G^2 + 2 b_G N_P \label{NG_eq} \\
\dot{N}_A & = & -2r_A N_A^2 + 2 b_A N_P - s_A N_A \label{NA_eq}\\
\dot{N}_P & = & g_P - b_A N_P + r_A N_A^2 - b_G N_P + r_G N_G^2 - e_P N_P \label{NP_eq}
\eea

We consider now the steady-state solution. Equation~\eqref{NG_eq}  gives
\be\label{NG_sol}
N_G = \sqrt{b_G N_P/r_G}
\ee
and the two terms before the last one in Eq.~\eqref{NP_eq} cancel out. Then we are left with the system
\bea
0 & = & -2r_A N_A^2 + 2 b_A N_P - s_A N_A \label{ssna} \\
0 & = & g_P - b_A N_P + r_A N_A^2 -e_P N_P \label{ssnp}
\eea
We can solve Eq.~\eqref{ssna} for $N_A$ in terms of $N_P$ to find
\be
N_A = \sqrt{\left(\frac{s_A}{4r_A}\right)^2+\frac{b_A}{r_A}N_P} - \frac{s_A}{4r_A}
\ee
and substituting into Eq.~\eqref{ssnp} we get
\be
0 = g_P - e_P N_P +\frac{s_A^2}{8r_A}\left(1-\sqrt{1+\frac{16 r_A b_A}{s_A^2}N_P}\right)
\ee
We are interested in finding $N_P$ as the pair-breaking rate for the phonon traps is varied (by changing the coverage). An approximate solution valid under the conditions
\be\label{NPcond}
\frac{s_A^2}{8g_P r_A} \ll 1 \, , \qquad \frac{b_A}{e_P} \gg \frac{s_A^2}{8g_P r_A}
\ee
has the form
\be
N_P = \frac{g_P}{e_P}\left[ 1 + \frac{s_A^2}{8g_P r_A} \frac{b_A}{e_P}-
\sqrt{2\frac{s_A^2}{8g_P r_A} \frac{b_A}{e_P}+\left(\frac{s_A^2}{8g_P r_A} \frac{b_A}{e_P}\right)^2} \right] \label{NPsol}
\ee

The first condition in \eref{NPcond} is the same weak scattering condition assumed in Ref.~\cite{valenti2018interplay}, while, under the second one, \eref{NPsol} reduces to the equation of $N_P$ reported in Ref.~\cite{valenti2018interplay}. Also, while this solution is strictly speaking invalid for $b_A \to 0$ (i.e., in the absence of phonon traps), it gives nonetheless the correct leading order result $N_P \approx g_P/e_P$ (the perturbative solution is $N_P\simeq g_P/(e_P + b_A) \approx g_P/e_P (1-b_A/e_P)$; the two solution agree at next-to-leading order at the crossover point $b_A/e_P \sim s_A^2/4g_Pr_A$). We will use \eref{NPsol} to calculate $N_P$ both in the absence and presence of phonon traps.

To proceed further, we need to know how the parameters depend on the size of the traps. To this end, we note that \eref{NA_eq}, for example, should be written for the (areal) density $x_A$ of QPs, rather than the total number:
\be
\dot{x}_A = -2\tilde{r}_A x_A^2 + 2 \tilde{b}_A x_P - \tilde{s}_A x_A
\ee
where parameters with tilde depend on material properties, but not on geometry. Using now the definition $x_\alpha = N_\alpha/A_\alpha$, we rewrite the above equation as:
\be
\dot{N}_A = -2\frac{\tilde{r}_A}{A_A}N_A^2 + 2\tilde{b}_A \frac{A_A}{A_P} N_P - \tilde{s}_A N_A,
\ee
where the area $A_P$ for phonons is the total area of the chip. We can repeat this step for the other two equations \rref{NG_eq} and \rref{NP_eq} [for the latter, note that $r_\alpha$ and $b_\alpha$ are non-zero only in the areas coverd by material $\alpha$], and we obtain the following relations for the parameters:
\be\label{pargeomdep}
g_P = \tilde{g}_P A_P\,, \quad b_\alpha =\tilde{b}_\alpha \frac{A_\alpha}{A_P}\,, \quad r_\alpha = \frac{\tilde{r}_\alpha}{A_\alpha}\, , \quad s_A = \tilde{s}_A\, , \quad e_P = \tilde{e}_P
\ee

We can now rewrite \eref{NPsol} in terms of parameters with tilde as
\be
\frac{N_P}{A_P} = \frac{\tilde{g}_P}{\tilde{e}_P}\left[ 1 +  (\beta F)^2-
\sqrt{2(\beta F)^2 +(\beta F)^4} \right] \, , \quad \beta^2 = \frac{\tilde{s}_A^2}{8\tilde{g}_P \tilde{r}_A} \frac{\tilde{b}_A}{\tilde{e}_P} \, , \quad  F = \frac{A_A}{A_P} \label{NP_fin}
\ee

Going now back to \eref{Qeq}, using \eref{NG_sol} we have
\be\label{QNP}
\frac{1}{Q_i} = c \sqrt{\frac{\tilde{b}_G}{\tilde{r}_G}\frac{N_P}{A_P}}
\ee
and using \eref{NP_fin}, we see that inverse $Q$-factor can be fitted using only two parameters: an overall dissipation rate $ 1/Q_{i,0} =c\sqrt{\tilde{b}_G\tilde{g}_P/\tilde{r}_G\tilde{e}_P}$, and the phenomenological factor $\beta$ (cf. Eq. \eqref{NP_fin}): 
\be
    \frac{1}{Q_i} = \frac{1}{Q_{i,0}} \sqrt{1 +  (\beta F)^2- \sqrt{2(\beta F)^2 +(\beta F)^4}}.
\ee
In the main text we show the result of such a fit where $\beta=9$ for all curves, giving $Q_{i,0}=8.3\times10^{4}$, $1.9\times10^{4}$, and $2\times10^{4}$ for A, B, and C, respectively.

\subsection{QP burst rate}

In the previous parts we have tacitly assumed that there is a steady-state population of phonons of energy somewhat larger (but not much larger) than $2\Delta_G$; this is qualitatively consistent with the scattering being ``weak'', as the gap $\Delta_G$ is only a couple of times as big as $\Delta_A$, and the QP scattering rate $s_A$ is suppressed for energy close to the gap $\Delta_A$ (see discussion in Ref.~\cite{valenti2018interplay} and references therein). This type of phonons, generating QPs just above the gap $\Delta_G$, contribute to limiting quality factor and to generation/recombination noise. For bursts, on the other hands, we assume that they are due to rarer phonons of much higher energy ($\gg 2\Delta_G$). Due to their higher energy, the QPs that they generate (in both $G$ and $A$) scatter at much higher rate, so the downconversion is much faster. In a simple model, we therefore expect that the rate at which such a high energy phonon disappear from the substrate has three contribution: it can escape from the substrate to the Cu waveguide holder with a rate $e_P^H$ (we use superscript $H$ to denote rates pertaining to these high energy phonons; due to the energy dependence of the various scattering and pair breaking mechanisms, these rates likely differ from the rates of the lower energy phonons considered above); or they can break pairs in the resonator or in the traps with rates $b_G^H$ and $b_A^H$, respectively. We neglect the unlikely possibility that the generated QPs recombine right away by emitting again a high-energy phonon. So the probability $P_H(t)$ of having a high-energy phonon at time $t$ obeys the equation
\be
\dot{P}_H (t) = - (e_P^H+b_A^H+b_G^H) P_H(t).
\ee
Assuming that the phonon is generated at time $t=0$, so that $P_H(0)=1$, the solution is then simply
\be
P_H(t) = e^{-(e_P^H+b_A^H+b_G^H)t}
\ee
Conversely, such a phonon is absorbed in the resonator at a rate $b_G^H$; assuming that once the phonon is absorbed, it starts a burst, the probability $P_B(t)$ of a burst then obeys the equation 
\be
\dot{P}_B(t) = b_G^H P_H(t)
\ee
with the initial condition $P_B(0)=0$. The solution is then
\be
P_B(t) = \frac{b_G^H}{e_P^H+b_A^H+b_G^H}\left(1-e^{-(e_P^H+b_A^H+b_G^H)t}\right)
\ee
The average probability of a burst is then
\be
\langle P_B \rangle = \lim_{T\to\infty}\frac{1}{T}\int_0^T P_B(t) \, dt = \frac{b_G^H}{e_P^H+b_A^H+b_G^H}
\ee
The burst rate $\Gamma_B$ is then the rate $g_H$ at which high-energy phonons are generated times the average probability that such a phonon causes a burst:
\be
\Gamma_B = g_H \frac{b_G^H}{e_P^H+b_A^H+b_G^H}
\ee
Assuming that rates depend on areas as in \eref{pargeomdep}, we find
\be \label{gamma_intermed}
\Gamma_B = \frac{g_H}{1+\frac{\tilde{b}_A^H A_P}{\tilde{b}_G^H A_G}\left(F+\frac{\tilde{e}^H_P}{\tilde{b}^H_A}\right)} 
\ee
Note that since $A_P \sim 1\,$cm$^2$ and we roughly expect $\tilde{b}^H_G \sim \tilde{b}^H_A$, then as order of magnitude $\tilde{b}^H_A A_P/\tilde{b}^H_G \sim 1\,$cm$^2$. Then, since $A_G < 10^{-3}\,$cm$^2$, we have $\tilde{b}^H_A A_P/\tilde{b}^H_G A_G < 10^{-3}$, and so long as $\tilde{e}^H_P/\tilde{b}^H_A \gg 10^{-3}$, we can neglect unity in the denominator and write
\be \label{gammab_eq_a}
\Gamma_B \simeq \frac{g_H \tilde{b}_G^H A_G/\tilde{b}_A^H A_P}{F+\tilde{e}_P^H/\tilde{b}_A^H} \equiv \Gamma_0 \frac{ \Lambda}{F+\Lambda},
\ee
where $\Lambda = \tilde{e}^H_P/\tilde{b}^H_A$ indicates the phonon relaxation ratio, i.e. the rate at which high energy phonons escape to the substrate scaled to the rate at which they break CPs in aluminum, and $\Gamma_0$ indicates the QP burst rate for the case with no traps. Note that $\Gamma_0 \sim 1/\tilde{e}_P^H$: a better phonon thermalization is always beneficial in reducing the QP burst rate. In the main text we have fitted Eq.~\eqref{gammab_eq_a} to the measured data using $\Lambda = 0.18$ for all resonators, which is indeed large compared to $10^{-3}$ (as required in the approximation going from Eq.~\eqref{gamma_intermed} to Eq.~\eqref{gammab_eq_a}), obtaining $\Gamma_0 = 7.7$, $5.6$, and $2.3 \times 10^{-2}$~$\text{s}^{-1}$ for resonators A, B and C, respectively.

\newpage 
\section{Resonator parameters} \label{app_circlefit}

\begin{figure*}[h!]
\begin{center}
\def\svgwidth{1 \textwidth}  
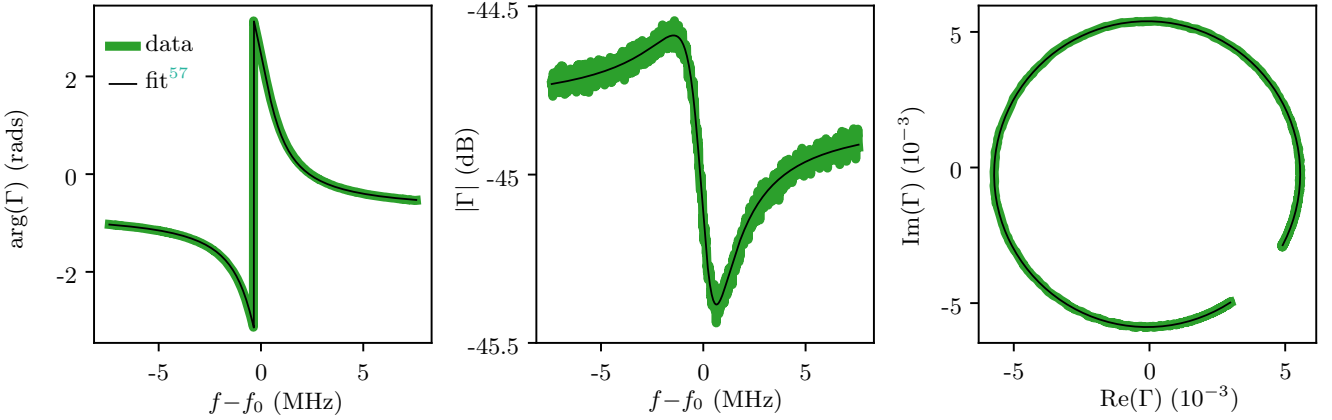
\caption{Reflection coefficients for resonator C, $F=19\%$. The panels from left to right show the phase, amplitude, and complex plane response, respectively.}\label{fit}
\end{center}
\end{figure*}

\begin{figure*}[h!]
\begin{center}
\def\svgwidth{1 \textwidth}  
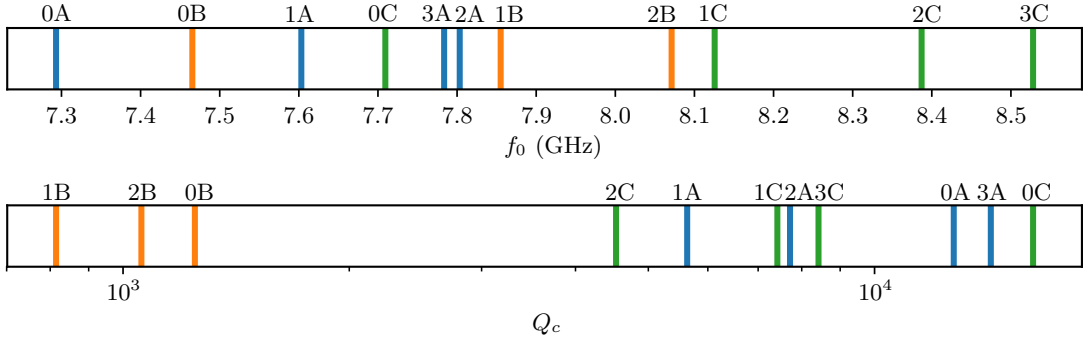
\end{center}
\caption{\textbf{a)} Resonant frequencies of resonators A, B, and C, shown in blue, orange, and green, respectively. Numbers 0 to 3 represent $F=0$, $8.5$, $19$ and  $34\%$, respectively. \textbf{b)} Coupling quality factors (same coloring/numbering of panel \textbf{a}).}\label{fit_par}
\end{figure*}

We extract the resonant frequency $f_0$, the internal quality factor $Q_i$ and coupling quality factor $Q_c$ of the 11 resonators using the method described in Ref.~\cite{shahid2011reflection}. An example of a fitted resonator reflection coefficient, measured at a strong drive before bifurcation, is shown in Fig.~\ref{fit}. We show the extracted $f_0$ and $Q_c$ for all resonators in Fig.~\ref{fit_par}. 
\vspace{10 mm}
\vspace{10 mm}

\vspace{10 mm}
\vspace{10 mm}

The resonant frequency shows an increasing trend for resonators labeled from 0 to 3 in Fig.~\ref{fit_par}. This trend corresponds to different resonator positions across the sapphire wafer. The frequency trend is also reflected in a gradient in the DC resistivity measured across the wafer, from $6\;\text{m}\Omega\; $cm at label 0 to  $4\;\text{m}\Omega\; $cm at label 3, which can be explained by a gradient in oxygen pressure during grAl deposition. Decreased normal state resistivity implies lower kinetic inductance, hence increased $f_0$. We identify resonators of type B by their lower $Q_c$, due to both their stronger dipole moment and central position in the waveguide providing the largest coupling to the electric field. Resonators of type A and C are differentiated via their resonance frequency, estimated with finite element simulations.

\newpage 
\section{Internal quality factor as a function of circulating photons}\label{app_qinbar}

\begin{figure*}[h!]
\begin{center}
\def\svgwidth{1 \textwidth}  
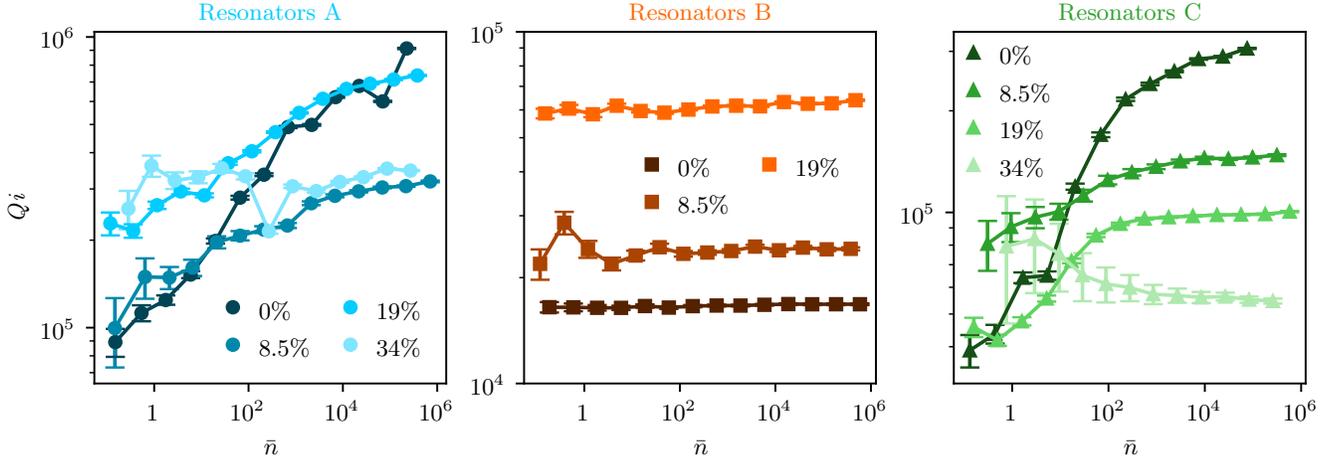
\caption{Internal quality factor as a function of average photon number. }\label{Qin}
\end{center}
\end{figure*}

We show the measured dependence of the internal quality factor $Q_i$ as a function of the number of drive photons $\bar{n}$ in Fig.~\ref{Qin}. 

\vspace{10 mm}
\vspace{10 mm}

\newpage 
\section{Noise spectral density as a function of readout photon number}\label{app_nsdpow}

\begin{figure*}[h!]
\begin{center}
\def\svgwidth{1 \textwidth}  
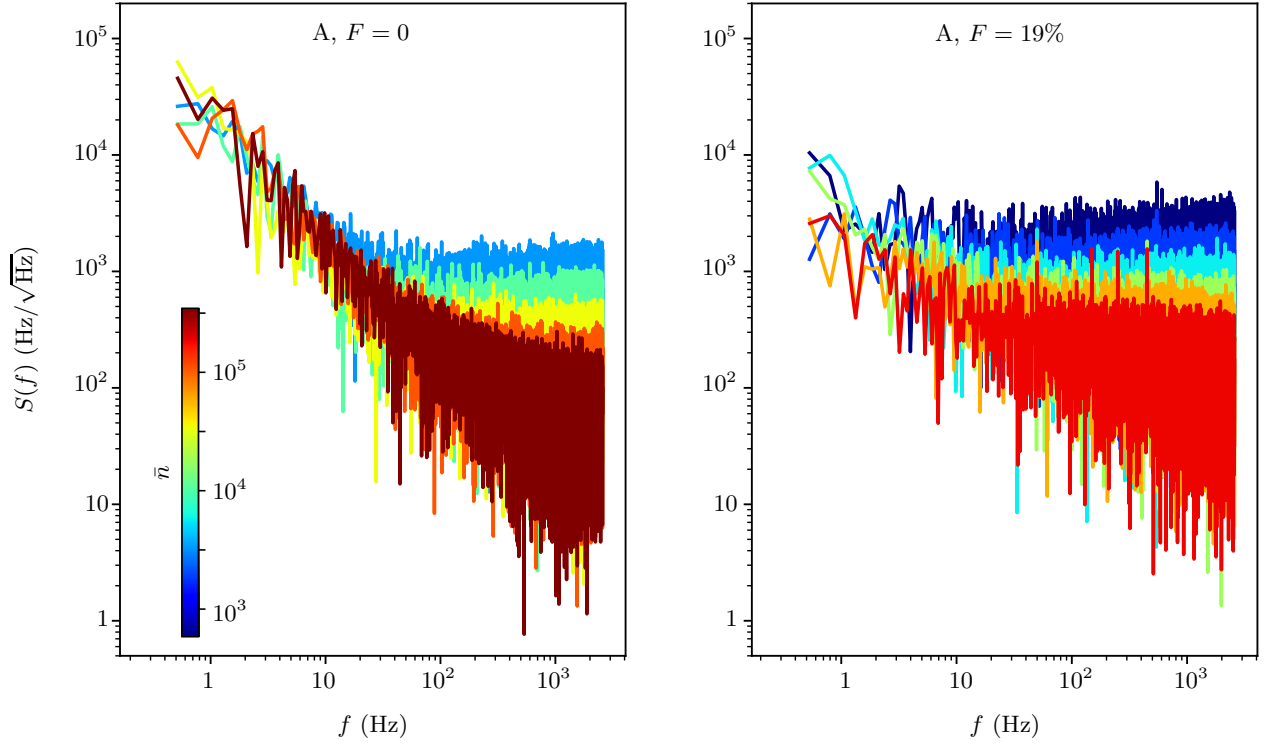
\caption{Noise spectral density as a function of readout photon number for resonators A, $F=0$ and A, $F=19\%$.}\label{nsdnbar}
\end{center}
\end{figure*}

We show the resonator noise spectral density dependence on the number of readout photons, $\bar{n}$, in Fig.~\ref{nsdnbar}. At higher $\bar{n}$, the improved microwave signal-to-noise ratio lowers the white noise floor. The $1/f$ portion of the spectra is not affected. 

\vspace{10 mm}
\vspace{10 mm}

\newpage 
\section{Quasiparticle burst rates over days}\label{app_days}

\begin{figure*}[h!]
\begin{center}
\vspace{3 mm}
\def\svgwidth{1 \textwidth}  
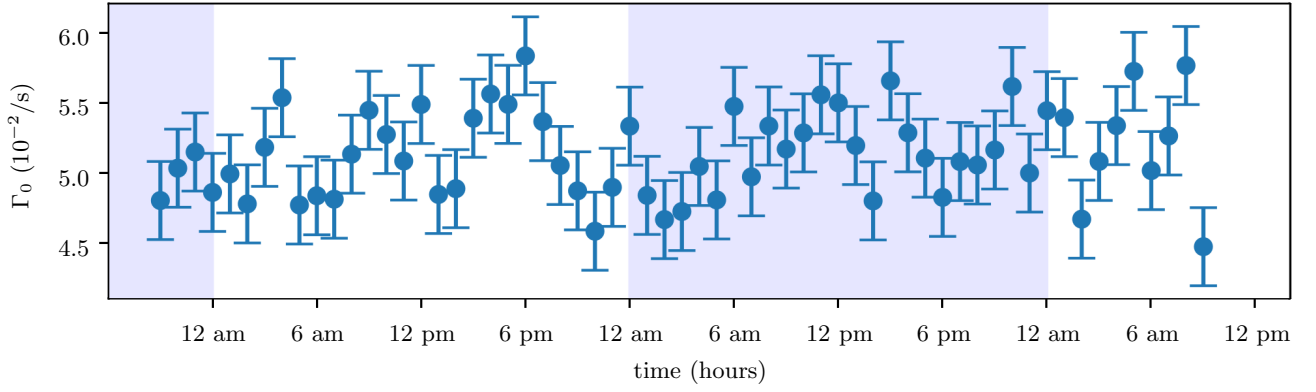
\caption{Measured QP burst rate $\Gamma_0$ over approximately $60$ hours for a typical grAl resonator. Days are shown as alternating light blue bands.}\label{ratedays}
\end{center}
\end{figure*}

In Fig.~\ref{ratedays}, we show the rate of QP bursts measured over approximately $60$ hours in a previous experiment, in the same setup and for a grAl resonator with the same size and kinetic inductance as resonator A. There is no appreciable drift over the total acquisition time, nor between daytime and nighttime. 

\vspace{10 mm}
\vspace{10 mm}

\newpage 
\section{Frequency evolution over thirty minutes}\label{app_spectrograms}

\begin{figure*}[h!]
\vspace{3 mm}
\begin{center}
\def\svgwidth{1 \textwidth}  
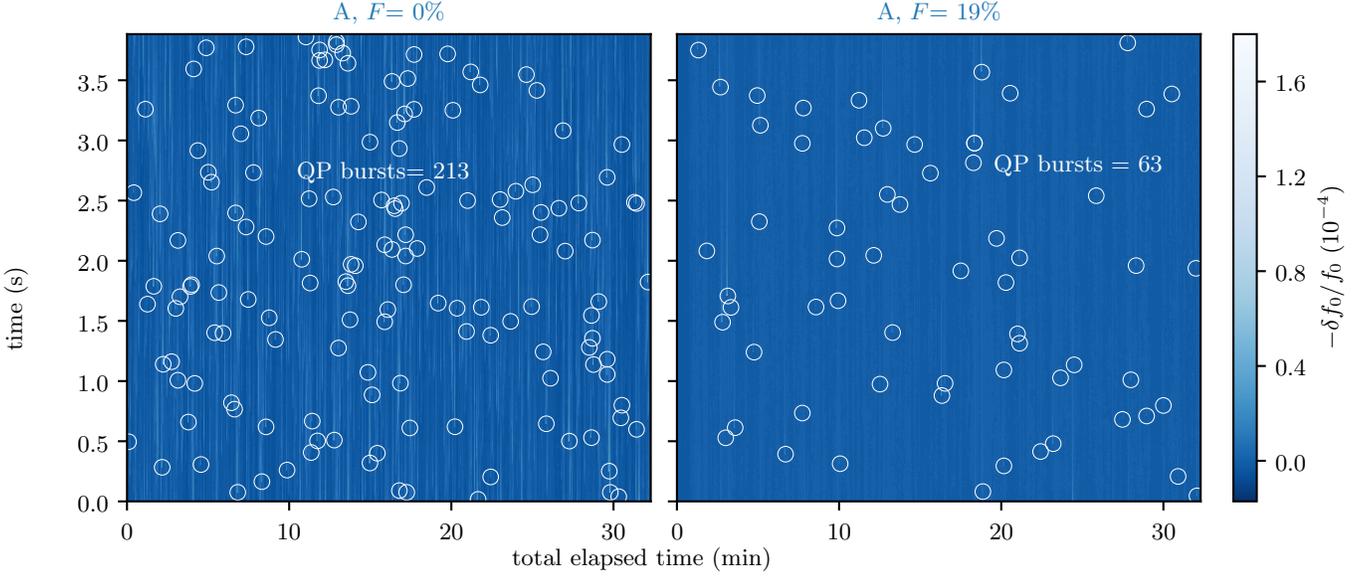
\caption{Typical 30 minutes long time traces for resonators A, $F=0$  (left panel) and $19\%$ (right panel). Quasiparticle bursts are marked with a white circle.}\label{longtt}
\end{center}
\end{figure*}

We measure the time evolution of the resonant frequency of each individual resonator for 10 hours. In Fig.~\ref{longtt} we show randomly selected, 30 minutes long time traces chosen for resonators A, $F=0$ and $19\%$. The resonator surrounded by phonon traps with $F=19\%$ shows a decrease in both fluctuations (cf. Fig.~\ref{fig_parameters}c and d in the main text) and measured QP bursts (cf. Fig.~\ref{fig_events} in the main text).

\vspace{10 mm}
\vspace{10 mm}

\end{document}